# Dicoogle Framework for Medical Imaging Teaching and Research

Rui Lebre[1,2] [a], Eduardo Pinho[1], Jorge Miguel Silva[1] and Carlos Costa[1]
[1]*Institute of Electronics and Informatics Engineering of Aveiro, University of Aveiro, Portugal*
[2]*Faculty of Computer Science, University of A Coruña, Spain*
{*ruilebre, eduardopinho, jorge.miguel.ferreira.silva, carlos.costa*}*@ua.pt*

Abstract: One of the most noticeable trends in healthcare over the last years is the continuous growth of data volume produced and its heterogeneity. In the medical imaging field, the evolution of digital systems is supported by the PACS concept and the DICOM standard. These technologies are deeply grounded in medical laboratories, supporting the production and providing healthcare practitioners with the ability to set up collaborative work environments with researchers and academia to study and improve healthcare practice. However, the complexity of those systems and protocols makes difficult and time-consuming to prototype new ideas or develop applied research, even for skilled users with training in those environments. Dicoogle emerges as a reference tool to achieve those objectives through a set of resources aggregated in the form of a learning pack. It is an open-source PACS archive that, on the one hand, provides a comprehensive view of the PACS and DICOM technologies and, on the other hand, provides the user with tools to easily expand its core functionalities. This paper describes the Dicoogle framework, with particular emphasis in its Learning Pack package, the resources available and the impact of the platform in research and academia. It starts by presenting an overview of its architectural concept, the most recent research backed up by Dicoogle, some remarks obtained from its use in teaching, and worldwide usage statistics of the software. Moreover, a comparison between the Dicoogle platform and the most popular open-source PACS in the market is presented.

## 1 INTRODUCTION

In an information society, data is a cornerstone for scientific progress. This is not an exception in healthcare, wherein the usage of digital medical imaging has become predominant and fundamental in clinical practice [1, 2]. The development of new modalities and the continuous improvements in digital medical imaging systems has led to the increase of data produced in institutions. The concept of Picture Archiving and Communication System (PACS) comprises the many technologies for acquisition, archiving, distribution and visualization of digital images using a computer network for diagnosis and revision in dedicated stations [3, 4].

PACS is deeply linked with the Digital Imaging and Communications in Medicine (DICOM) standard, which defines data formats, storage organization and communication protocols of digital medical imaging [5]. As an international standard, it guarantees the interoperability between equipment and information systems [6] in a PACS, by ensuring normalized data formats and processes among each component. This environment has subsequently led to a steadfast research and conception of new systems, namely of imaging acquisition modalities [7] and tools for data analysis. Moreover, as the number of medical imaging data produced is ever increasing, the means of searching for relevant information in a medical imaging archive poses as a challenge that, in recent years, has been tackled with new ways of searching for information [8, 9, 10]. Hence, the increased complexity of a modern PACS makes proper training on how these systems work and interact a necessity.

In the last decade, the University of Aveiro Bioinformatics research group has been developing the Dicoogle project[2] as a tool to achieve this goal, with its current use having been expanded to the industry, academia and research labs. Placing it as a platform for both users and developers, Dicoogle provides a comprehensive view of the operation of PACS and DICOM standard technologies, as well as tools to expand these technologies. In this article, we expose Dicoogle Learning Pack framework for research and education, describing how its resources teach the necessary understandings of medical imaging informatics with the use of a modern PACS archive and helping to overcome the difficulty of usage and navigation of the system. As such, this document targets students and researchers, but also raises its equivalent potential use by the industry. Furthermore, Dicoogle is compared with other open source PACS systems. In parallel, we compare the solutions according to a defined set of requirements.

---

[a] https://orcid.org/0000-0002-3230-0732

[2]Official website:

## 2 RELATED WORK

Until recently, PACS systems were not used as academic tools because existing systems did not offer easy-to-use functionality or the ability to augment them in a decoupled fashion. However, the fast development of medical imaging systems and networks (evidently expanded to beyond the traditional concept of radiography), and the increasing legal requirements of digital information in developed and developing countries have made education an essential requirement for all physicians and radiologists. As such, the creation and usage of these systems seemed inevitable [11].

Small steps have been made towards this goal over the last decade. For instance, several authors have tried to use PACS for the creation of electronic teaching files (ETF), which comprise clinical-case medical images, pertinent information about the clinical details, and diagnosis. Lim et al. [12, 13] developed an ETF server which retrieves DICOM images from a PACS and enables users to create teaching files, while Wilkinson et al. [14] integrated digital data supplied by an institutional PACS system into teaching files.

On the other hand, there are also attempts to integrate teaching modules with PACS. For instance, Sinha et al. [15] developed an interactive teaching module, linked to a PACS database, for teaching radiology. Likewise, Towbin et al. [13] developed a computer-based radiology simulator that mimicked the local hospital PACS to create a learning tool to help prepare first-year residents for being on call.

In terms of research, there were also some notable concerns, since the traditional clinical PACS archive was insufficiently flexible for the requirements of research. This is mainly due to the existence of two completely separate workflows: the clinical workflow and the research workflow. In the clinical scenario, data is transferred daily to the institutional PACS, normally from local imaging devices, and viewed using proprietary software. Whereas in research, data enter the system from a rich variety of sources and both the original data and the processed results usually reside in a personal space, without any common organizational structure or centrally available history of the processing methods performed [16]. Hence, due to the existence of these two distinct workflows, the need for a *research oriented PACS archive* arose.

In this line, Doran et al. [16] described a prototype research PACS framework, which is based on XNAT in association with tools of data selection and archiving. Also, Haak et al. [17] integrated DICOM data into electronic data capture systems (EDCS) regarding the workflow in clinical trials and a PACS archiving node for research.

Jodogne et al. [18, 19] described an open-source standalone DICOM store for healthcare and medical research, capable of deploying multiple instances in a hospital network or a standalone computer, providing a RESTful API, and can be bundled with an embedded Web interface. On the same line, Warnock et al. [20] describes dcm4chee Archive, a modular designed PACS built out of the previously tools and frameworks of JDicom. Woodbridge et al. [21] shows MRIdb as a self-contained image database designed to manage MRI datasets. Eichelberg et al. describes in [22] DCMTK, a collection of libraries and applications implementing parts of the DICOM standard and Stott et al. [23] created a open-source archiving and management of ultrasound images.

Furthermore, other PACS have been identified: Xebra[3], CDMedic PACS Web [4], JSVdicom server[5], DicomServer@Medical Connections[6], ClearCanvas[7], K-PACS[8] and ConquestDICOM[9].

## 3 LEARNING PACK FRAMEWORK

### 3.1 Dicoogle Concept

Dicoogle is an open-source PACS archive software that replaces the traditional database model with an extensible indexing and retrieval framework [24, 25] and provides easy expansion of functionalities through the use of plug-ins. It was designed to ac-commodate automatic information extraction, indexing, and storage of all meta-data detected in medical images, without re-engineering or reconfiguration requirements, thus overcoming the limitations of traditional DICOM query services [26]. By presenting the technical assets for plug-in development such as a Software Development Kit (SDK), developers are free to expand the archive independently and non-exclusively, without changes to the core platform.

---

[3]**Xebra**: sourceforge.net/projects/xebra
[4]**CDMedic**: sourceforge.net/projects/cdmedicpacsweb
[5]**JSVdicom**: jvsmicroscope.uta.fi
[6]**DicomServer@Medical Connections**: www.dicomserver.co.uk
[7]**ClearCanvas**: www.clearcanvas.ca
[8]**K-PACS**: www.k-pacs.net
[9]**ConquestDICOM**: ingenium.home.xs4all.nl

This extensible architecture of Dicoogle has enabled its use in research and the healthcare industry, as many use cases can be fulfilled in the same deployment [7, 27]. This is very relevant nowadays, given the need to improve, monitor and measure the efficiency of medical imaging systems, as well as to extract knowledge from the produced medical images, including healthcare quality indicators. As such, Dicoogle has been successfully used as a base platform for DICOM data mining [28, 29] and as an archive for exploiting multimodal information retrieval methods [9].

## 3.2 Framework Architecture

Figure 1 presents the architecture of Dicoogle. The design of the framework is separated in five distinct categories according to its functionality: *Index*, *Query*, *Storage*, *Web Services*, and *Web UI Components*. Each category is tied to a specific application programming interface (API), which is implemented by plug-ins. Through these interfaces, the plug-ins provide operations that are orchestrated by the core Dicoogle platform.

The life cycle of each plug-in is controlled by Dicoogle's core, that during the application startup scans the plug-in directory and identifies the plug-ins which are then loaded according to their directives present in the configuration file.

Dicoogle core sees the plug-ins as completely independent from each other, being only accessible via the respective interfaces. In the case of the need to share states between plug-ins, the dependencies are solved inside the `PluginSet`, a class that represents a set of plug-ins and serves as an entry point and management entity to the overall structure of the extension. Here, plug-ins of several categories are aggregated into one functionally consistent unit, thus simplifying the development and deployment process.

The Dicoogle SDK provides the necessary software components for interconnecting plug-ins with the core system. Although it is not presented as a center-piece, this SDK simplifies development by establishing the APIs that the plug-ins must abide to, as well as implementing interfaces with the system for accessing indexed and non-indexed data, dispatching tasks, fetching settings and logging the software's functions.

The usual line of thought for creating a new extension to Dicoogle can be summarized into the following points:

1. The developer has an idea for a feature that is tightly related to the PACS archive.

2. They create a new project from a bare-bones project (in practice, a copy from the sample) a base plug-in set (Listing 1).

3. Specific plug-ins are implemented based on the given requirements (Listing 2 presents an example of the methods to be implemented in an index plug-in).

 (a) If Dicoogle is going to contain additional sources of information, it is usual to create an *indexer* to process that information from incoming medical images, and a *query* provider to expose that information to the system.

 (b) When making a plug-in for remote storage (such as to a cloud service), a *storage* plug-in can be created for the intended storage provider. The same type of plug-in can also be used for constructing pre-processing pipelines, which can transform inbound files and delegate the storage proper to another plug-in.

 (c) *Web services* are the most flexible, as they enable the web application, and other client programs, to fetch new kinds of information from the server and perform new operations.

4. For extending the main user interface, *web user interface* plug-ins should be considered instead (Section 3.3).

Further instructions for helping the Dicoogle plug-in development are available online at Dicoogle Learning Pack [10].

---

[10]Available at: https://bioinformatics-ua.github.io/dicoogle-learning-pack/

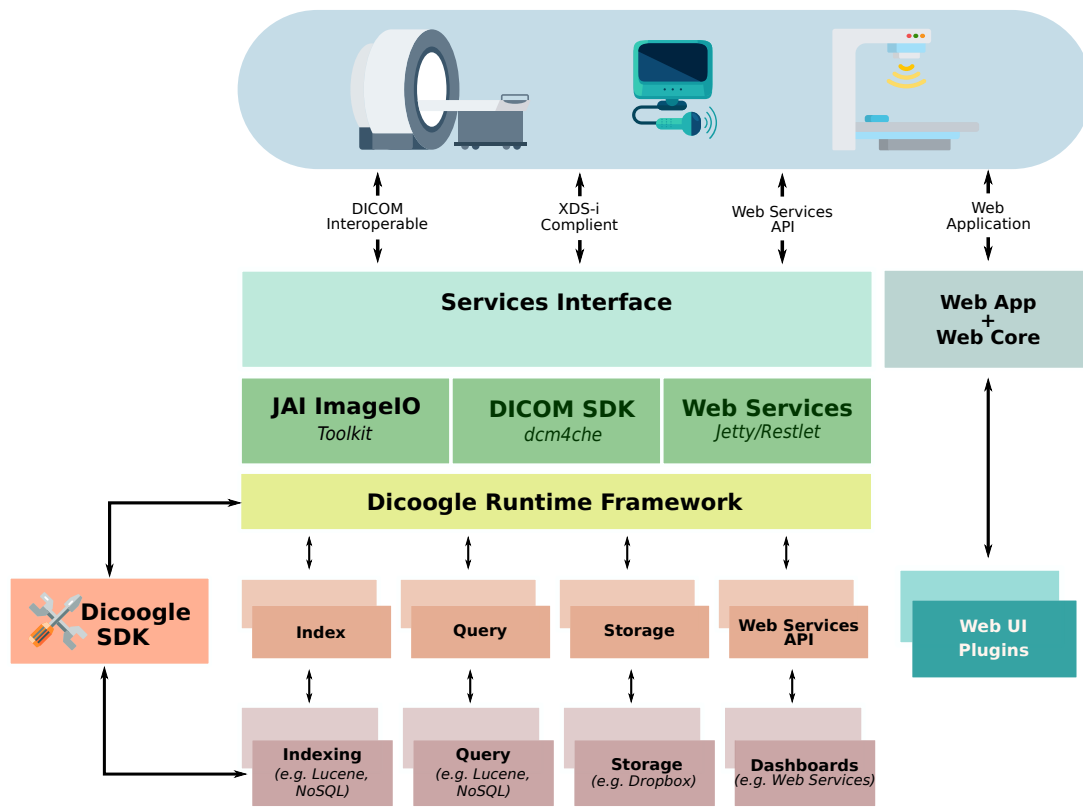

Figure 1: Dicoogle Architecture

Listing 1: Plug-in development example - plug-in declaration on PluginSet. Available at Dicoogle Learning Pack - Developing Plug-ins.

```java
@PluginImplementation
public class MyPluginSet implements
PluginSet {
    // use slf4j for logging purposes
    private static final Logger logger =
    LoggerFactory.getLogger(MyPluginSet.class);

    // You can list each of our plugins as an
     attribute to the plugin set
    private final MyQueryProvider query;

    // Additional resources may be added here.
    private ConfigurationHolder settings;

    public MyPluginSet() throws IOException
    {
        logger.info("Initializing My Plugin
        Set");

        // construct all plugins here
        this.query = new MyQueryProvider();

        logger.info("My Plugin Set is ready");
    }

    @Override
    public Collection<QueryInterface>
    getQueryPlugins() {
        return
        Collections.singleton((QueryInterface)
        this.query);
    }

    @Override
    public String getName() {
        return "mine";
    }

    // ... implement the remaining methods
}
```

Listing 2: Index plug-in example. Available at Dicoogle Learning Pack - Developing plug-ins

```
/**
 Index  Interface  Plugin .  Indexers  analyze  documents
 for  performing  queries .  They  may  index  documents  by
 DICOM  metadata  for  instance ,  but  other  document
 processing  procedures  may  be  involved .
*/
public interface IndexerInterface extends
DicooglePlugin {

    /**
     Indexes  the  file  path  to  the  database .  Indexing
     procedures  are  asynchronous ,  and  will  return
     immediately  after  the  call .  The  outcome  is  a
     report  that  can  be  retrieved  from  the  given
     task  as  a  future .

     @param  file  directory  or  file  to  index
     @return  a  representation  of  the  asynchronous
     indexing  task
     */
    Task<Report> index(StorageInputStream
    file, Object ... parameters);

    /**
     Indexes  multiple  file  paths  to  the  database .
     Indexing  procedures  are  asynchronous ,  and  will
     return  immediately  after  the  call .  The  outcomes
     are  aggregated  into  a  single  report  and  can  be
     retrieved  from  the  given  task  as  a  future .

     @param  files  a  collection  of  directories  and / or
     files  to  index
     @return  a  representation  of  the  asynchronous
     indexing  task
     */
    Task<Report>
    index(Iterable<StorageInputStream> files,
            Object ... parameters);

    /**
     Checks  whether  the  file  in  the  given  path  can
     be  indexed  by  this  indexer .  The  indexer  should
     verify  if  the  file  holds  compatible  content
     (e.g.  a  DICOM  file ).  If  this  method  returns
     false ,  the  file  will  not  be  indexed .

     @param  path  a  URI  to  the  file  to  check
     @return  whether  the  indexer  can  handle  the  file
     at  the  given  path
     */
    boolean handles(URI path);

    /**
     Removes  the  indexed  file  at  the  given  path  from
     the  database .

     @param  path  the  URI  of  the  document
     @return  whether  it  was  successfully  deleted
     from  the  database
     */
    boolean unindex(URI path);
}
```

## 3.3 Service Interface

Dicoogle provides a modern single-page web user interface, which can be delivered to local and remote users without a previous installation process on the client machine.

To accommodate new use cases from both ends of the software stack, Dicoogle also includes a dynamic pluggable web component architecture. Rather than manually including new menus and actions to the web application, the Dicoogle *Web Core* establishes a backbone for web UI component retrieval and rendering in runtime at the browser. This enables the exposure of new features on the main web application without rebuilding the source code.

To achieve this in practice, three components were developed:

1. A set of web services in the Dicoogle web server for providing the web-based plug-ins and their meta-information;

2. The Web Core component, a JavaScript library that runs in the user's browser for retrieving and exposing the extra user interfaces;

3. A project scaffolding tool to facilitate the creation of web UI plug-ins.

Each user interface plug-in is scoped to a particular view of the application, into divisions called slots. Slots describe the kind of interface that a plug-in for the same slot is meant to portray. For example, the `settings` slot is available in the management section. When observing a list of results, a `result-entry` slot is created for each item, so as to perform a particular operation over an item, such as visualize the image in a separate service). In contrast, plug-ins of the type `result-batch` are scoped to the entire list of results, allowing the exposure of new forms for data visualization, exporting, and manipulation.

The interactions of the Web Core with these slots and the main web application are represented in Figure 2. Once the user is logged in, the application fetches the list of web user interface plug-ins available for that user, using standard asynchronous requests. The actual components, specified as Javascript modules, are subsequently loaded and rendered to the indicated slot on the page. The Web Core architecture relies on standard HTML5 web components to augment slot elements for this functionality.

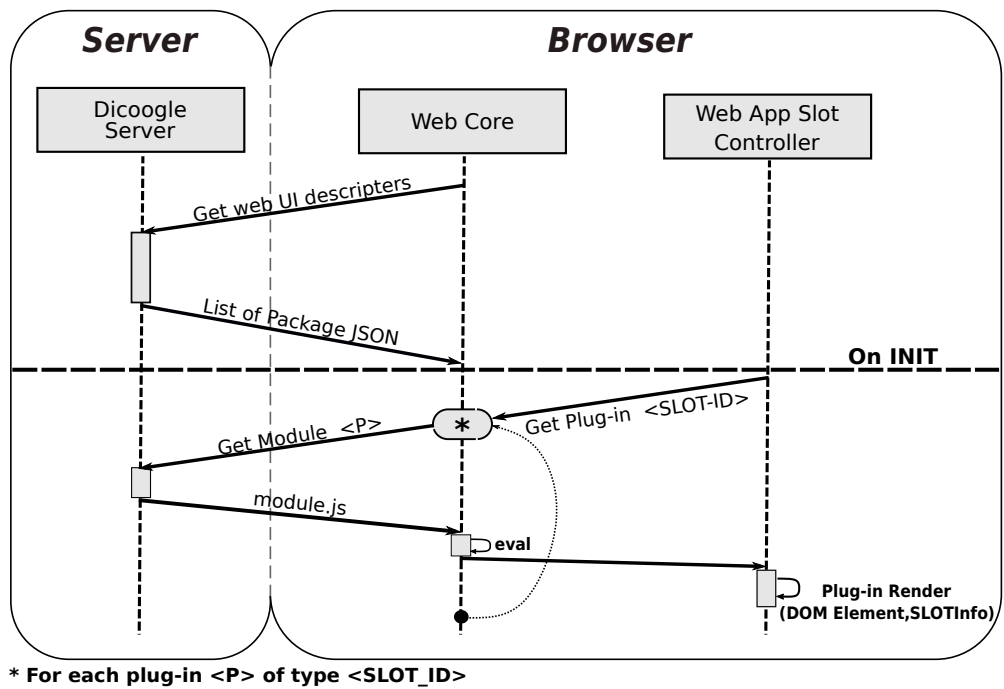

Figure 2: Dicoogle Web Core sequence diagram

## 3.4 Learning Pack Resources

Medical imaging informatics is a major subject and teaching offer is becoming increasingly frequent. The notable increase in higher education degrees and courses that combine radiology concepts with medical imaging informatics is a direct consequence of their impact on healthcare.

Therefore, teaching how to use an open-source PACS archive in a medical imaging informatics context to students is an essential counterpart to the development of the actual software. With this in mind, the Dicoogle Learning Pack was created to teach users with no prior or deep knowledge to tackle new advances of the field in an academic environment, backed by the Dicoogle archive. The resources were designed to be as simple and objective as possible to be a relevant tool to minimize and smooth the open-source PACS and DICOM standard learning curve.

The Dicoogle Learning Pack[11] is available as an open-access static site hosted on GitHub Pages. The website itself is open-source, and users of Dicoogle are invited to read and provide their feedback through GitHub's issue tracker. This approach takes advantage of freely available resources for open-source projects while making them accessible to researchers and students alike. Moreover, the opening of issues and active participation of the community in discussion, allows the improvement of both the Dicoogle Learning Pack and Dicoogle itself, meeting the expectation of the players overtime. The Learning Pack provides also documentation, configurations, and code examples, guiding the user to, for instance:

- Install Dicoogle for the first time on a local machine with any operating system;
- Configure indexers, services and plugins of any kind;
- Index a data set and perform searches over the indexed data;
- Develop plugins for the Dicoogle back-end, using Dicoogle SDK, in which a background in Java is assumed and following a specific workflow as shown in Figure 3;
- Develop web user interface plugins for the Dicoogle web application, in which a basic background in Javascript programming knowledge and notions of HTML are assumed;
- As a more advanced topic, the build the main Dicoogle application is introduced, divided by building of the back-end and front-end;
- Debug the Dicoogle main application source-code and web-app, as well as the plugins source-code.

---

[11] https://bioinformatics-ua.github.io/dicoogle-learning-pack

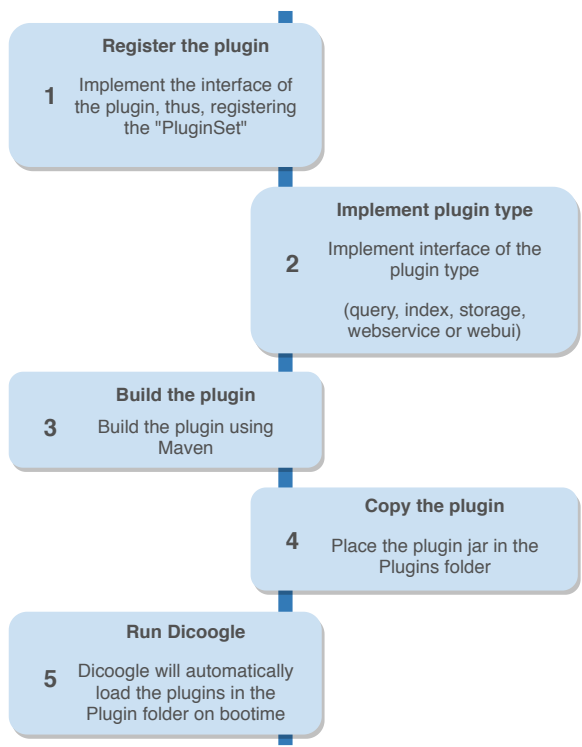

Figure 3: Workflow of the development of a Dicoogle Plugin

### 3.4.1 Index and Query Plugins

Being an archive, features as indexing and retrieving information are crucial in Dicoogle. Dicoogle can have its technologies for storing DICOM meta-data or even processed data extended. That way, Dicoogle is not bound to a single implementation. This is achieved with the Query and Index plugin pair, which usually go hand-in-hand:

- The indexer plugins are triggered when incoming requests for indexing objects reach Dicoogle and processes it for future retrieval;
- Query plugins interpret queries from the user and return lists of results based on the provided parameters.

### 3.4.2 Storage Plugins

Data storage is one of the main goals of a PACS archive. As the technology evolves and more patient data is acquired, medical imaging storage requirements have become increasingly demanding. The server's local file system has become inadequate in some use cases. Delegating storage to cloud services, for instance, have been thoroughly discussed in recent years [30, 31].

Storage plugins are responsible to deal with the place where files can be read and stored, and sometimes also define how these files are stored and read. The storage plugins are triggered every time the PACS archive receives a C-STORE request and handle the persistence of the DICOM object. The Dicoogle Storage API serves the following two purposes:

- creating an abstraction over the underlying storage technology, thus being able to use and evaluate different sources of data storage (e.g. cloud storage services such as Amazon S3) and different forms of persistence (e.g. using a document-oriented database instead of a file system). With this common API, DICOM object reading and storing becomes possible, regardless of the underlying technology;
- enhancing the storage and retrieval procedures with certain algorithms, such as for anonymization, compression, and encryption.

### 3.4.3 Web Services Plugins

Web services are one of the most flexible ways of expanding Dicoogle with new features. Currently, there are two ways to achieve this:

- Jetty Servlets can be created and registered using a plugin of type JettyPluginInterface. In this type of plugins, one can create their own servlets, then attach them into a handler list. The servlet API is a versatile mean of creating services in Java server applications. A servlet class should inherit from the HttpServlet class. It may then override one or more of the HTTP request method handlers (doGet, doPost, doPut, and so on).
- Rest Service plugins consider a subset of the REST framework API, allowing developers to create and attach simple server resources. The RESTlet API is more straightforward than Jetty Servlets, albeit more limited in some cases. The REST services implemented are directly described by method prototypes and annotations instead of a fixed set of methods.

### 3.4.4 Web UI Plugins

Dicoogle web user interface plugins are pluggable components that can be integrated in the web application. The developers are mandated to specify a type of plugin, often with the "slot-id" property. This type definition specifies the way that the web plugins are attached to the application. The Web UI plugins can be be found in four types:

- *menu*: menu plugins are used to extend the main menu. A new entry is added to the side bar, and the component is created when the user navigates to that entry.

- *result-option*: result option plugins are used to provide advanced operations to a result entry. If the user activates "Advanced Options" in the search results view, these plugins will be attached into a new column, one for each visible result entry.

- *result-batch*: result batch plugins are used to provide advanced operations over an existing list of results. These plugins attach a button which will pop-up a division below the search result view.

- *settings*: settings plugins can be used to provide additional management information and control. These plugins are attached to the "Plugins & Services" tab in the Management menu.

## 4 USAGE SCENARIOS

Dicoogle has been used for fulfilling a variety of use cases in medical imaging informatics since its inception. For instance, in the last decade, more than ten PhD students made use of Dicoogle framework in their research works. Below, it is presented the experimental use of the Dicoogle Learning Pack for teaching, followed by two of the latest lines of research where Dicoogle served as the backbone.

### 4.1 Teaching Networks and Services in Imaging

At the University of Aveiro, a particular subject is simultaneously attended by students of radiology (physicians) and students of engineering. On a yearly basis, the Electronics, Telecommunications, and Informatics department gives Master's and Doctoral degree students in computer science, as well as Master's students in medical imaging technologies, the opportunity to enroll on the subject of Networks and Services in Imaging. The class teaches an average of 30 students every year. Prior to their enrollment, students of Medical Imaging Technology have limited knowledge of computer science. On the other hand, computer engineering students are not familiar with either the DICOM standard or the PACS concepts. The teaching of systems and networks in medical imaging comprises two complementary perspectives:

- Experts in *radiology* may understand the main purposes of a PACS archive and know how to use them for research and clinical use, including some understanding of the technical challenges overcome by these systems;

- From a perspective of *computer engineering and software development*, students are given the necessary background of PACS systems and networks, including the DICOM standard, in order to develop end-user solutions.

As such, the objective of this joint course is to introduce students to the generic concept of PACS and the DICOM Standard, addressing the main medical imaging modalities, as well as quality management and control issues in PACS environments and radiology information systems.

In order to learn and consolidate these concepts, students have to execute a three-month long development projects related to the PACS/DICOM universe, for which most students have used Dicoogle and the Learning Pack. For instance, the projects "Workflow Management in Distributed Medical Imaging Environments" and "3D Defacing - Visual Anatomic Anonymization", which culminated into the production of scientific publications [32, 33]. At the end of the academic year, it was sent an anonymous form to students to evaluate these tools as learning resources.

Overall, the results obtained from the forms indicated that, regarding Dicoogle:

- 63.7% found Dicoogle very useful as a learning resource;
- 54.6% thought it was easy to set up;
- 42.8% found the process of creating plug-ins easy;

On the other hand, regarding Dicoogle Learning Pack:

- 74.5% considered the Dicoogle Learning Pack was well structured;
- 87.5% thought it was useful in setting up and general usage of Dicoogle;
- 57.2% felt that the learning pack made the development of plug-ins significantly easier;

This results provides some insight that these resources are useful as a learning resource and as a tool of knowledge consolidation regarding concepts like PACS and the DICOM Standard. For those who used Dicoogle in their subject final project, it was deemed a vital tool for the creation of new projects.

### 4.2 Dicoogle as a framework in Research

Nowadays, the exploration and analytics of medical data from the medical imaging laboratories is a com-pelling component in the healthcare institutions businesses. The huge amount of data can reveal statistics and help to forecast measures to improve the quality of healthcare services, for instance. Godinho et al. [34] described a business intelligence framework taking advantage of the Dicoogle data mining tools over DICOM.

Pedrosa et al. [35] presented a collaborative platform of the SCREEN-DR project. The SCREEN-DR project addresses the automatic detection of diabetic retinopathy (DR) by developing computer-aided methods. The authors deployed, on top of Dicoogle open-source PACS, a platform to collaboratively annotate visually and textually the DICOM images dataset. The ultimate goal is to use those annotations to the conception of machine learning methods to aid the screening process.

Medical image search capabilities are boosted by the combination of textual and visual features. Yet, there is still a lack of support in the most of the imaging archives, as they only index the available meta-data. In [36], the authors described an architecture for automatic labeling, extracting visual information from DICOM images. This work extended Dicoogle capabilities by adding support for automated content discovery with multi-modal querying that, otherwise, would only be possible with a CAD framework incorporation.

With the technological assistance of the Dicoogle platform, a reversible de-identification mechanism was developed by Silva et al. [7]. In this use case, standard medical imaging objects are fully de-identified, including meta-data and pixel data. At the same time, it provides a reversible de-identification mechanism that retains search capabilities from the original data. The goal was to deploy this solution in a collaborative platform where data is anonymized when shared with the community, but still searchable for data custodians or authorized entities.

During the last decades, Digital Pathology has been arising as a new branch of pathology [37]. Digital Pathology is the aggregation of hardware and software designed to substitute traditional devices, like microscopes. The acquisition of the images (whole-slide images) is performed by whole-slide scanners. Those obtained images can have a resolution of several gigapixels and, therefore, consume large amounts of space when stored digitally. The developed system by Godinho et al. [38] allows the deployment of a performant and fully DICOM compliant viewer of whole-slide images using Dicoogle as the base PACS archive. Dicoogle was also extended to support the creation of DICOM WSI image pyramid. The solution provides a tiling engine and the viewer uses standard DICOM Web services implemented as Dicoogle plug-ins.

# 5 RESULTS

In order to evaluate the impact of Dicoogle among the scientific community, a survey was conducted to discover currently available open-source PACS software. The research was based on the past review papers [39, 40] and by the use of the query "open-source PACS" in Google Scholar search engine. The list of the gathered 12 solutions encompassed Dicoogle [24, 25, 41], Orthanc [18, 19], MRIdb [21], DCMTK (Offis) [22], dcm4chee [20], Xebra, OSPACS, CDMedic PACS Web, JSVdicom server, DicomServer@Medical Connections, ClearCanvas, K-PACS and ConquestDICOM.

First, we compared the different open-source PACS in terms of usability and ease of development. To this end, we verified if the source code of the PACS system was still available as well as the date of their last update. We discarded all PACS that no longer had source code available or that did not receive an update for the past few years. The remaining PACS were: DCMTK (Offis), Orthanc, dcm4chee, MRIdb and Dicoogle.

In this extended analysis, some parameters were selected, such as supported platforms, user guide, the existence of integrated web viewer, plug-in based extensibility, development framework available, documentation available, REST API, DICOMWeb support and digital pathology support. The results were standardized and shown in Table 1.

Dicoogle and Orthanc have shown more coverage in terms of satisfying criteria, being the strongest deployed systems both on academic and clinical production environments.

In fact, Dicoogle started to attain worldwide visibility since the release of the official Dicoogle website in 2014. Since then, the site has accumulated more than 10.000 downloads of Dicoogle. However, the solution can be downloaded directly from GitHub and the absolute number of downloads in this platform is not available. As shown in Figure 4, Dicoogle is used worldwide as a research and teaching platform, with the largest number of users being from the United States, Europe, Brazil and India.

The website was rebuilt from scratch at the end of 2017 and the Learning Pack was made available to the community, attaining a greater user reach and a stronger impact in the community with about 175 downloads per month during 2018 and almost 200 downloads per month until November 2019.

Table 1: Analysis of the most updated open-source PACS in October 2018.

| | Last Update | Sourcecode Availability | Supported Platforms | User Guide | Web Viewer | Plug-in Extensible | Development Framework | Docu-mentation | REST API | DICOMWeb Support | DICOM Pathology Support |
|---|---|---|---|---|---|---|---|---|---|---|---|
| **MRIdb** | 07/08/2018 | GitHub | Linux | | ✓ | | | ✓ | | ✓ | |
| **DCMTK (Offis)** | 01/10/2019 | GitHub | Windows Linux Mac | | | | | ✓ | ✓ | | |
| **Orthanc** | 10/10/2019 | GitHub Bitbucket | Windows Linux Mac | ✓ | ✓ | ✓ | ✓ | ✓ | ✓ | ✓ | ✓ |
| **dcm4chee** | 10/10/2019 | GitHub | Windows Linux Mac | | ✓ | ✓ | ✓ | ✓ | ✓ | ✓ | |
| **Dicoogle** | 12/09/2019 | GitHub | Windows Linux Mac | ✓ | ✓ | ✓ | ✓ | ✓ | ✓ | ✓ | ✓ |

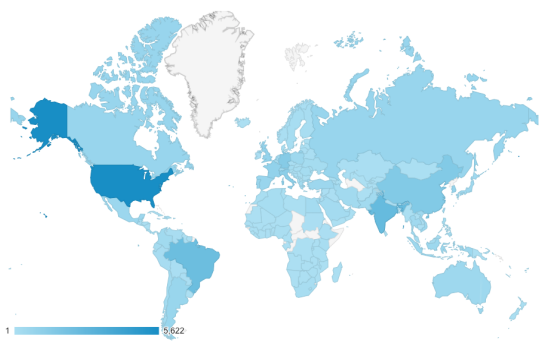

Figure 4: Distribution of Dicoogle usage worldwide, based on the number of accesses to the official website.

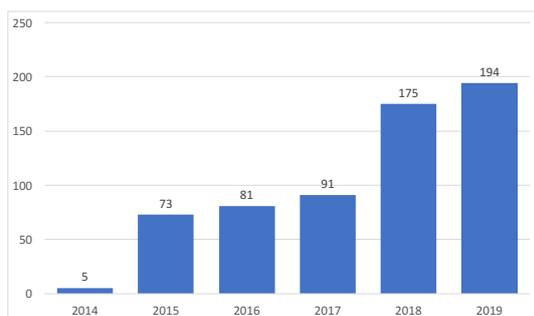

Figure 5: Average number of Dicoogle downloads per month in each year.

As shown in Figure 5, the number of monthly downloads of this framework has been increasing over the years, suggesting an increase in the use and adoption of this technology. This increase makes us believe that this tool, together with the learning resources that promote the usage and navigation of this system, can have a big impact in research and education.

## 6 CONCLUSIONS

This paper presents the Dicoogle project and its Learning Pack as a resource designed to support research and academia. Dicoogle is a web based open-source PACS archive with expandable capabilities, providing an insight over its learning resources that enable researchers to quickly set up and use this platform. Some representative uses of this tool have been show-cased in distinct application areas, supported by different technologies that were appended to Dicoogle as plug-ins. Finally, a comparison between the most popular open-source PACS in the market and Dicoogle is presented, as well as evidences of Dicoogle's worldwide growth in terms of platform usage and its usefulness in education. Hence, it is concluded that Dicoogle is a relevant tool that, together with the appropriate learning resources, can have a big impact in research and education of medical imaging informatics.

## ACKNOWLEDGEMENTS

This work has received support from the ERDF European Regional Development Fund through the Operational Programme for Competitiveness and Internationalization, COMPETE 2020 Programme, and by National Funds through the FCT, Fundação para a Ciência e a Tecnologia within the project PTDC/EEI-ESS/6815/2014. Jorge Miguel Silva was funded by FCT Portugal under the grant SFRH/BD/141851/2018.

## References

[1] Dhiraj Manohar Dhane et al. "Fuzzy spectral clustering for automated delineation of chronic

[1] wound region using digital images". In: *Computers in biology and medicine* 89 (2017), pp. 551–560.

[2] Reza Moradi Rad et al. "A hybrid approach for multiple blastomeres identification in early human embryo images". In: *Computers in biology and medicine* 101 (2018), pp. 100–111.

[3] Keith J. Dreyer et al., eds. *PACS*. New York: Springer-Verlag, 2006. ISBN: 0-387-26010-2. DOI: 10.1007/0-387-31070-3. URL: http://link.springer.com/10.1007/0-387-31070-3.

[4] H K Huang. *PACS and Imaging Informatics: Basic Principles and Applications*. 2nd. Wiley, 2010. ISBN: 9780470560518. URL: http://books.google.pt/books?id=Pjjkyae%7B%5C_%7D55oC.

[5] Oleg S. Pianykh. *Digital Imaging and Communications in Medicine (DICOM) - A Practical Introduction and Survival Guide*. Springer, 2012, p. 417. ISBN: 9783642108495. DOI: 10.2967/jnumed.109.064592. URL: http://books.google.pt/books?id=tFgSRghUzGsC.

[6] Peter Mildenberger, Marco Eichelberg, and Eric Martin. "Introduction to the DICOM standard". In: *European Radiology* 12.4 (Apr. 2002), pp. 920–927. ISSN: 0938-7994. DOI: 10.1007/s003300101100. URL: http://link.springer.com/10.1007/s003300101100.

[7] Jorge Miguel Silva et al. "Controlled searching in reversibly de-identified medical imaging archives". In: *Journal of Biomedical Informatics* 77 (2018), pp. 81–90. ISSN: 1532-0464. DOI: https://doi.org/10.1016/j.jbi.2017.12.002. URL: https://www.sciencedirect.com/science/article/pii/S1532046417302721.

[8] Eduardo Pinho et al. "A Multimodal Search Engine for Medical Imaging Studies". In: *Journal of Digital Imaging* 30.1 (Feb. 2017), pp. 39–48. ISSN: 1618-727X. DOI: 10.1007/s10278-016-9903-z. URL: https://doi.org/10.1007/s10278-016-9903-z.

[9] E. Pinho and C. Costa. "Extensible Architecture for Multimodal Information Retrieval in Medical Imaging Archives". In: *2016 12th International Conference on Signal-Image Technology Internet-Based Systems (SITIS)*. Nov. 2016, pp. 316–322. DOI: 10.1109/SITIS.2016.58.

[10] John Puentes et al. "Integrated multimedia electronic patient record and graph-based image information for cerebral tumors". In: *Computers in biology and medicine* 38.4 (2008), pp. 425–437.

[11] Robert Sigal. "PACS as an e-academic tool". In: *International Congress Series* 1281 (2005), pp. 900–904. ISSN: 0531-5131. DOI: https://doi.org/10.1016/j.ics.2005.03.240. URL: http://www.sciencedirect.com/science/article/pii/S053151310500498X.

[12] C C Tchoyoson Lim et al. "Medical Image Resource Center–making electronic teaching files from PACS". In: *Journal of Digital Imaging* 16.4 (2003), pp. 331–336. ISSN: 1618-727X. DOI: 10.1007/s10278-003-1660-0. URL: https://doi.org/10.1007/s10278-003-1660-0.

[13] Cct Lim and Gl Yang. "Electronic teaching files and continuing professional development in radiology." eng. In: *Biomedical imaging and intervention journal* 2.2 (Apr. 2006), e5. ISSN: 1823-5530 (Print). DOI: 10.2349/biij.2.2.e5.

[14] Luke E Wilkinson and Sam R Gledhill. "An Integrated Approach to a Teaching File Linked to PACS". In: *Journal of Digital Imaging* 20.4 (2007), pp. 402–410. ISSN: 1618-727X. DOI: 10.1007/s10278-006-1045-2. URL: https://doi.org/10.1007/s10278-006-1045-2.

[15] S Sinha et al. "A PACS-based interactive teaching module for radiologic sciences." eng. In: *AJR. American journal of roentgenology* 159.1 (July 1992), pp. 199–205. ISSN: 0361-803X (Print). DOI: 10.2214/ajr.159.1.1609698.

[16] Simon J Doran et al. "Informatics in Radiology: Development of a Research PACS for Analysis of Functional Imaging Data in Clinical Research and Clinical Trials". In: *RadioGraphics* 32.7 (2012), pp. 2135–2150. DOI: 10.1148/rg.327115138. URL: https://doi.org/10.1148/rg.327115138.

[17] Daniel Haak et al. "DICOM for clinical research: PACS-integrated electronic data capture in multi-center trials". In: *Journal of digital imaging* 28.5 (2015), pp. 558–566.

[18] S Jodogne et al. "Orthanc - A lightweight, restful DICOM server for healthcare and medical research". In: *2013 IEEE 10th International Symposium on Biomedical Imaging*.

2013, pp. 190–193. DOI: 10.1109/ISBI.2013.6556444.

[19] Sébastien Jodogne. "The Orthanc Ecosystem for Medical Imaging". In: *Journal of digital imaging* 31.3 (2018), pp. 341–352.

[20] Max J Warnock et al. "Benefits of using the DCM4CHE DICOM archive". In: *Journal of Digital Imaging* 20.1 (2007), pp. 125–129.

[21] Mark Woodbridge, Gianlorenzo Fagiolo, and Declan P O'Regan. "MRIdb: medical image management for biobank research". In: *Journal of digital imaging* 26.5 (2013), pp. 886–890.

[22] Marco Eichelberg et al. "Ten years of medical imaging standardization and prototypical implementation: the DICOM standard and the OFFIS DICOM toolkit (DCMTK)". In: *Medical Imaging 2004: PACS and Imaging Informatics*. Vol. 5371. International Society for Optics and Photonics. 2004, pp. 57–69.

[23] Will Stott et al. "OSPACS: Ultrasound image management system". In: *Source Code for Biology and Medicine* 3.1 (2008), p. 11.

[24] Frederico Valente et al. "Anatomy of an Extensible Open Source PACS". In: *Journal of Digital Imaging* 29.3 (June 2016), pp. 284–296. ISSN: 0897-1889. DOI: 10.1007/s10278-015-9834-0.

[25] Frederico Valente, Carlos Costa, and Augusto Silva. "Dicoogle, a PACS featuring profiled content based image retrieval." In: *PloS one* 8.5 (2013), e61888. ISSN: 1932-6203. DOI: 10.1371/journal.pone.0061888. URL: http://www.ncbi.nlm.nih.gov/pubmed/23671578%20http://www.pubmedcentral.nih.gov/articlerender.fcgi?artid=PMC3646026.

[26] Carlos Costa et al. "Indexing and retrieving DICOM data in disperse and unstructured archives". In: *International Journal of Computer Assisted Radiology and Surgery* 4.1 (Jan. 2009), pp. 71–77. ISSN: 1861-6410. DOI: 10.1007/s11548-008-0269-7. URL: http://link.springer.com/10.1007/s11548-008-0269-7.

[27] Tiago Marques Godinho et al. "A Routing Mechanism for Cloud Outsourcing of Medical Imaging Repositories". In: *IEEE Journal of Biomedical and Health Informatics* 20.1 (Jan. 2016), pp. 367–375. DOI: 10.1109/JBHI.2014.2361633. URL: http://www.ncbi.nlm.nih.gov/pubmed/25343773%20http://www.ncbi.nlm.nih.gov/pubmed/25343773.

[28] Milton Santos et al. "DICOM Metadata Analysis for Population Characterization: A Feasibility Study". In: *Procedia Computer Science* 100 (2016). International Conference on ENTERprise Information Systems/International Conference on Project MANagement/International Conference on Health and Social Care Information Systems and Technologies, CENTERIS/ProjMAN / HCist 2016, pp. 355–361. ISSN: 1877-0509. DOI: https://doi.org/10.1016/j.procs.2016.09.169. URL: http://www.sciencedirect.com/science/article/pii/S1877050916323377.

[29] Milton Santos et al. "DICOM and Clinical Data Mining in a Small Hospital PACS: A Pilot Study". In: *ENTERprise Information Systems*. Ed. by Maria Manuela Cruz-Cunha et al. Berlin, Heidelberg: Springer Berlin Heidelberg, 2011, pp. 254–263. ISBN: 978-3-642-24352-3.

[30] James Philbin, Fred Prior, and Paul Nagy. "Will the next generation of PACS be sitting on a cloud?" In: *Journal of digital imaging* 24.2 (2011), pp. 179–183.

[31] Tim Rostrom and Chia-Chi Teng. "Secure communications for PACS in a cloud environment". In: *2011 Annual International Conference of the IEEE Engineering in Medicine and Biology Society*. IEEE. 2011, pp. 8219–8222.

[32] João Rafael Almeida et al. "Services Orchestration and Workflow Management in Distributed Medical Imaging Environments". In: *2018 IEEE 31st International Symposium on Computer-Based Medical Systems (CBMS)*. IEEE. 2018, pp. 170–175.

[33] Jorge Miguel Silva et al. "Face De-Identification Service for Neuroimaging Volumes". In: *2018 IEEE 31st International Symposium on Computer-Based Medical Systems (CBMS)*. IEEE. 2018, pp. 141–145.

[34] Tiago Marques Godinho et al. "ETL Framework for Real-Time Business Intelligence over Medical Imaging Repositories". In: *Journal of digital imaging* (2019), pp. 1–10.

[35] Micael Pedrosa et al. "SCREEN-DR: Collaborative platform for diabetic retinopathy". In: *International journal of medical informatics* 120 (2018), pp. 137–146.

[36] Eduardo Pinho and Carlos Costa. "Automated Anatomic Labeling Architecture for Content Discovery in Medical Imaging Repositories". In: *Journal of medical systems* 42.8 (2018), p. 145.

[37] Shaimaa Al-Janabi, André Huisman, and Paul J Van Diest. *Digital pathology: Current status and future perspectives*. July 2012. DOI: 10.1111/j.1365-2559.2011.03814.x. URL: http://doi.wiley.com/10.1111/j.1365-2559.2011.03814.x.

[38] Tiago Marques Godinho et al. "An efficient architecture to support digital pathology in standard medical imaging repositories". In: *Journal of Biomedical Informatics* 71 (2017), pp. 190–197. ISSN: 1532-0464. DOI: https://doi.org/10.1016/j.jbi.2017.06.009. URL: http://www.sciencedirect.com/science/article/pii/S1532046417301326.

[39] Kari Björn. "Evaluation of Open Source Medical Imaging Software: A Case Study on Health Technology Student Learning Experience". In: *Procedia Computer Science* 121 (2017), pp. 724–731.

[40] Stanisław Wideł, Andrzej Wideł, and Dominik Spinczyk. "Overview of available open source PACS frameworks". In: *Studia Informatica* 37.3A (2016), pp. 21–30.

[41] Carlos Costa et al. "Dicoogle - an Open Source Peer-to-Peer PACS". In: *Journal of Digital Imaging* 24.5 (Oct. 2011), pp. 848–856. ISSN: 1618-727X. DOI: 10.1007/s10278-010-9347-9. URL: https://doi.org/10.1007/s10278-010-9347-9.